\documentclass[12pt,preprint]{aastex}
\usepackage{epsfig}

\newcommand{\beq}{\begin{equation}}
\newcommand{\eeq}{\end{equation}}
\newcommand{\beqa}{\begin{eqnarray}}
\newcommand{\eeqa}{\end{eqnarray}}

\def\la{\mathrel{\mathpalette\fun <}}
\def\ga{\mathrel{\mathpalette\fun >}}
\def\fun#1#2{\lower3.6pt\vbox{\baselineskip0pt\lineskip.9pt
  \ialign{$\mathsurround=0pt#1\hfil##\hfil$\crcr#2\crcr\sim\crcr}}}

\begin{document} 

\title{The Ups and Downs of Baryon Oscillations} 
\author{Eric V.\ Linder} 
\affil{Physics Division, Lawrence Berkeley Laboratory, 
Berkeley, CA 94720}

%\nobreak 

%\vspace{0.2in} 

\begin{abstract} 
Baryon acoustic oscillations, measured through the patterned 
distribution of galaxies or other baryon tracing objects on very large 
($\ga100$ Mpc) scales, offer a possible geometric probe of cosmological 
distances.  Pluses and minuses in this approach's leverage for 
understanding dark energy are discussed, as are systematic uncertainties 
requiring further investigation.  Conclusions are that 1) BAO offer promise 
of a new avenue to distance measurements and further study is warranted, 
2) the measurements will need to attain $\sim1$\% accuracy (requiring 
a 10000 square degree spectroscopic survey) for their dark energy 
leverage to match that from supernovae, but do give complementary 
information at 2\% accuracy.  Because of the ties to the matter 
dominated era, BAO is not a replacement probe of dark energy, but a 
valuable complement. 
\vspace{-0.1in}
\end{abstract}

\section{Introduction} 

This paper provides a pedagogical introduction to baryon acoustic 
oscillations (BAO), accessible to readers not necessarily familiar with 
details of large scale structure in the universe.  In addition, it 
summarizes some of the current issues -- plus and minus -- with the 
use of BAO as a cosmological probe of the nature of dark energy.  
For more quantitative, technical discussions of these issues, see 
\citet{mwhite05}. 

{
The same year as the detection of the cosmic microwave background, the 
photon bath remnant from the hot, early universe, \citet{sakharov} predicted 
the presence of acoustic oscillations in a coupled baryonic matter 
distribution.  In his case, baryons were coupled to cold electrons 
rather than hot photons; \citet{peeblesyu} and \citet{sunyaev} 
pioneered the correct, hot 
case.  In the modern picture of these oscillations 
(see \citet{mwp} for a comprehensive technical treatment, and 
\citet{eisanim} for an exciting visual perspective), 
when the universe 
was hot enough for matter to be ionized, the photons and (charged) 
baryons were tightly coupled through electromagnetic forces.  This 
made the mean free path for the photons short compared to $ct$, the 
free streaming distance, and so the photons and baryons acted as a 
fluid medium, capable of supporting perturbations in the form of 
acoustic waves. 
}

The largest scale of the acoustic waves was set by the sound horizon. 
Due to the rapidity of the decoupling process after the universe 
recombined, and the lack of significant interactions thereafter, these 
largest scale wavelengths remain in close to their primordial state 
today.  Such oscillations take the form of patterns on this primordial 
sound horizon scale, and harmonics, in the spatial distribution of 
photons and baryons.  These acoustic waves in the photon 
number (or temperature) distribution were detected some 33 years 
after the CMB discovery.  The acoustic waves in the baryon spatial 
distribution were detected in 2005. 

The pattern in the CMB photons shows up as peaks and troughs of order 
unity deviation, making precision measurement of the angular scale of the 
primordial sound horizon possible with modern wide area surveys 
such as the WMAP satellite.  The theoretical derivation of the physical 
scale as a function of cosmology is straightforward due to the simple, 
well understood physics 
entering the photon-baryon coupling and decoupling, and the linear nature 
of the acoustic perturbations, due to fluctuation amplitudes of less than 
$10^{-4}$ in the 
number density, seeded (presumably) by early universe inflation. 
From the two elements of the measured angular scale 
and theoretical physical scale, one obtains the angular distance to 
the decoupling epoch. 

The baryon side of the effect is similar, but with some important 
differences.  For one thing, we do not detect baryons directly in the 
way we detect photons.  Instead we detect light emitted from processes 
involving baryons or electrons, or light affected in some way by the 
gravitational potential of mass in a structure (galaxy or cluster, say). 
While electrons should trace the baryon pattern well, and can be 
neglected in the mass effects (since a proton outweighs an electron by 
some 2000 times), other important components of mass exist besides 
baryons.  Indeed cold dark matter particles contribute six times 
more than baryons to the mass density of the universe.   Thus, the 
spatial pattern of oscillations traced by massive structures has been 
diluted relative to the primordial baryon acoustic oscillations.  
Furthermore, the baryons after decoupling found a ready made pattern 
of gravitational potentials, from the cold dark matter, waiting to 
influence them.  So the amplitude of the baryon acoustic oscillations 
is not of order one, like the photons, but rather $\la5\%$.  On 
the plus side, these oscillations are not relative to a $<10^{-4}$ 
base perturbation amplitude, but rather one of order $10^{-1}$, 
since the matter perturbations have been amplified by gravitational 
instability since the time of decoupling. 

Why then isn't it trivial to detect baryon acoustic oscillations, 
if their absolute amplitude is so much greater than the photon 
acoustic peaks?  Unfortunately, many more CMB photons are available 
to be detected -- $10^{15}$ pass through an outstretched hand each 
second, while there are fewer than $10^{10}$ galaxies in the entire 
visible universe.  Furthermore, there is much greater 
confusion of the baryon signal; many photons from the faint light 
given off by a distant galaxy must be detected before we even know 
there is a galaxy there tracing the primordial baryon distribution, 
and even more before we precisely know its three dimensional spatial 
location -- and that is for just one galaxy in the pattern. 
Still, this feat has been accomplished with the Sloan Digital Sky Survey 
(\citet{eis05}). 

Using the same theoretical physical horizon scale calculated for the 
CMB photons, and measuring the angular scale of the baryon acoustic 
oscillations at some redshift from a large galaxy redshift survey, 
their ratio provides the 
angular distance scale to that redshift.  This is often called a 
standard ruler test, in analogy to the supernova (SN) standard candle 
test.  Because the measured angle 
is the ratio of the detected angular scale to the physical horizon 
at decoupling, this distance measure is in some sense tied to the 
early universe rather than to the recent universe.  This will be 
important later (see \S\ref{sec.dist}).  Moreover, because one can 
measure a three dimensional 
pattern of galaxies (while only a two dimensional picture of the CMB 
sky), one can also derive a ratio corresponding to a radial 
distance, basically the Hubble parameter at some redshift. 

Thus baryon acoustic oscillations (BAO) offers a distance probe of the 
universe, and the expansion history and cosmological parameters including 
dark energy properties that enter the distance.  Because the galaxies 
(or other baryon sensitive objects) are used merely as markers of 
spatial position, BAO is a geometric test in the sense of not needing 
to know galaxy properties or masses (this is not absolutely true, as 
discussed in \S\ref{sec.sys}).  Only the wavelength of the 
oscillations follows simply from the primordial coupling to the CMB, 
not the amplitude, so BAO does not characterize the cosmic growth history. 
In this sense, this probe is in the same class as Type Ia supernovae. 

In the following text, we consider the positive and negative aspects 
of BAO as a cosmological probe, and the role it can play alone and 
in complementarity with other techniques.  We will see that, like 
every probe, corrections need to be applied.  We identify some systematic 
issues that need to be addressed for robust estimation of its power 
in constraining dark energy. 

\section{BAO are simple, SN are complicated {\it or} \\ 
SN are rich, BAO are meager?} 

We indicated above why BAO are so much harder to detect and characterize 
than CMB acoustic oscillations, and that after all the work of observing 
distant galaxies one ends with a single quantity for each galaxy -- its 
three dimensional position.  This is both a feature and a bug.  The use 
of galaxies as markers does not depend on the galaxy properties, other 
than as they influence the detection (selection effects).  SN 
are also used as geometric markers of the expansion history, but 
their distance measures depend on their luminosities, as well as 
intervening effects on the detected flux, such as dust.  Properties of 
the SN that influence the detected flux complicate their use as 
cosmological probes to the extent that we remain ignorant of such 
effects, i.e.\ how much they contribute systematic uncertainties. 

Since BAO depend on the galaxy positions, they appear much simpler. 
But systematic uncertainties arise here too.  For example, just as 
the SN flux must be properly translated to its emitted luminosity, 
the measured galaxy redshifts must be put in terms of their radial 
position.  This seems straightforward to accomplish, but at the level 
of precision required for a dark energy probe, corrections from 
redshift space to real space are not automatically trivial. 
We cannot a priori assume that 
although BAO appear to involve simple physics that their systematic 
uncertainties are negligible; we must compute them through theoretical 
calculations and compare them to simulations, and try the method 
out with real world data.  We discuss some of these uncertainties 
in the next section. 

In the presence of systematics, we must take seriously the section title 
question.  Indeed the 
dependence of BAO on just a single measured property offers the feature 
of simplicity, but is this a bug as well?  Each SN observed does not 
provide merely a single data point on the distance-redshift diagram, 
but rather a rich array of information that serves for crosschecks 
and systematics control.  A SN used in a cosmological survey has a 
complete flux history from shortly after explosion, through maximum 
light, and into the nebular phase, spanning over two months in the 
SN rest frame.  This is obtained in multiple passbands, covering the 
rest-frame visible light and possibly extending into the ultraviolet and 
near infrared.  An image of the SN relative to its host galaxy is 
part of the data, giving details on galaxy type and morphology, location 
of the SN in the core or outskirts, etc.  A spectrum of the SN provides 
detailed information on the physics, including through line velocities, 
shapes, and strengths.  This rich stream of data allows robust crosschecks 
on the use of the SN as a cosmological distance indicator.  No such 
crosschecks exist for the BAO method.  

So complexity, 
and simplicity, can be either a feature or a bug.  Neither should be 
automatically ruled out, but rather the use of all the data and the impact 
of remnant systematic uncertainties must be calculated in detail.  We 
do point out, however, that both methods have distinct advantages in 
being geometric methods, where the objects are markers only.  This also 
means that cosmologically unbiased selection of the best markers is allowed 
to provide the tightest constraints possible on cosmological parameters. 

\section{Systematic Uncertainties for Baryon Acoustic 
Oscillations} \label{sec.sys} 

While BAO indeed appear cleaner in physics basis and application than 
many other probes, this does not mean we should neglect careful 
scrutiny and computation to verify this.  Because of the short history 
of measurement of BAO, the list below of areas needing examination 
is likely incomplete. 

\subsection{Bias} \label{sec.bias} 

As mentioned above, while the primordial spatial pattern exists in 
baryons, we measure the light from assemblages of matter.  The relation 
between these quantities is referred to as the bias, and can in general 
be scale varying.  It seems likely that on the large scales of the BAO 
pattern, the bias will be smooth.  Preliminary simulation studies have 
been carried out by \citet{seoeis05} agreeing with this.  For a smooth 
``tilt'' to the matter power spectrum caused by bias, one can add 
fitting parameters, e.g.\ polynomial coefficients.  This can remove 
the effects of bias on measuring the oscillation scale, but it is not 
clear how much the additional parameters degrade or otherwise affect 
the scale estimation when trying to achieve percent level accuracy.  
Some simulations do appear to show 
an increasing shift in oscillation peak location as the bias increases, 
even after this correction. 

Another approach 
has been proposed by \citet{dolney}, where halo model bias parameters 
are fit by means of measuring higher order correlations plus the matter 
power spectrum.  So while definitive calculations need to be done, 
bias uncertainty is likely to be tractable and probably will not 
substantially interfere with the BAO method. 

\subsection{Nonlinear mode coupling} 

As pointed out in the Introduction, BAO actually have an advantage 
over the acoustic oscillations in the CMB in that the absolute amplitude 
of the matter fluctuations when measured in the recent universe is much 
higher than the photon density fluctuations.  However, this means that 
the simple, linear physics treatment as in the CMB does not transfer 
over with the same high degree of robustness -- one might say that 
the CMB is 99.99\% linear while the BAO are 90-99\% linear.  This is a 
nonneglible difference. 

The slight degree of nonlinearity causes coupling between modes, or 
scales, in the baryon spatial pattern.  This coupling increases as the 
matter fluctuations grow, becoming more significant as one approaches 
the more recent universe.  Such scale coupling smears the baryon 
acoustic oscillations, rendering them difficult to discern and possibly 
changing the scale.  One can avoid these effects by only looking at the 
longest wavelengths (where unfortunately sample variance, or the size 
of the survey, gives an increasingly large error), but then one 
restricts the number of oscillations available for measuring 
the standard ruler scale.  Commonly, one estimates that one to two peaks 
are visible for observations at $z\approx0.3$, three to four peaks 
for $z\approx1.5$, and six to seven peaks for $z\approx3$. 

\begin{figure}[!hbt]
\begin{center} 
\psfig{file=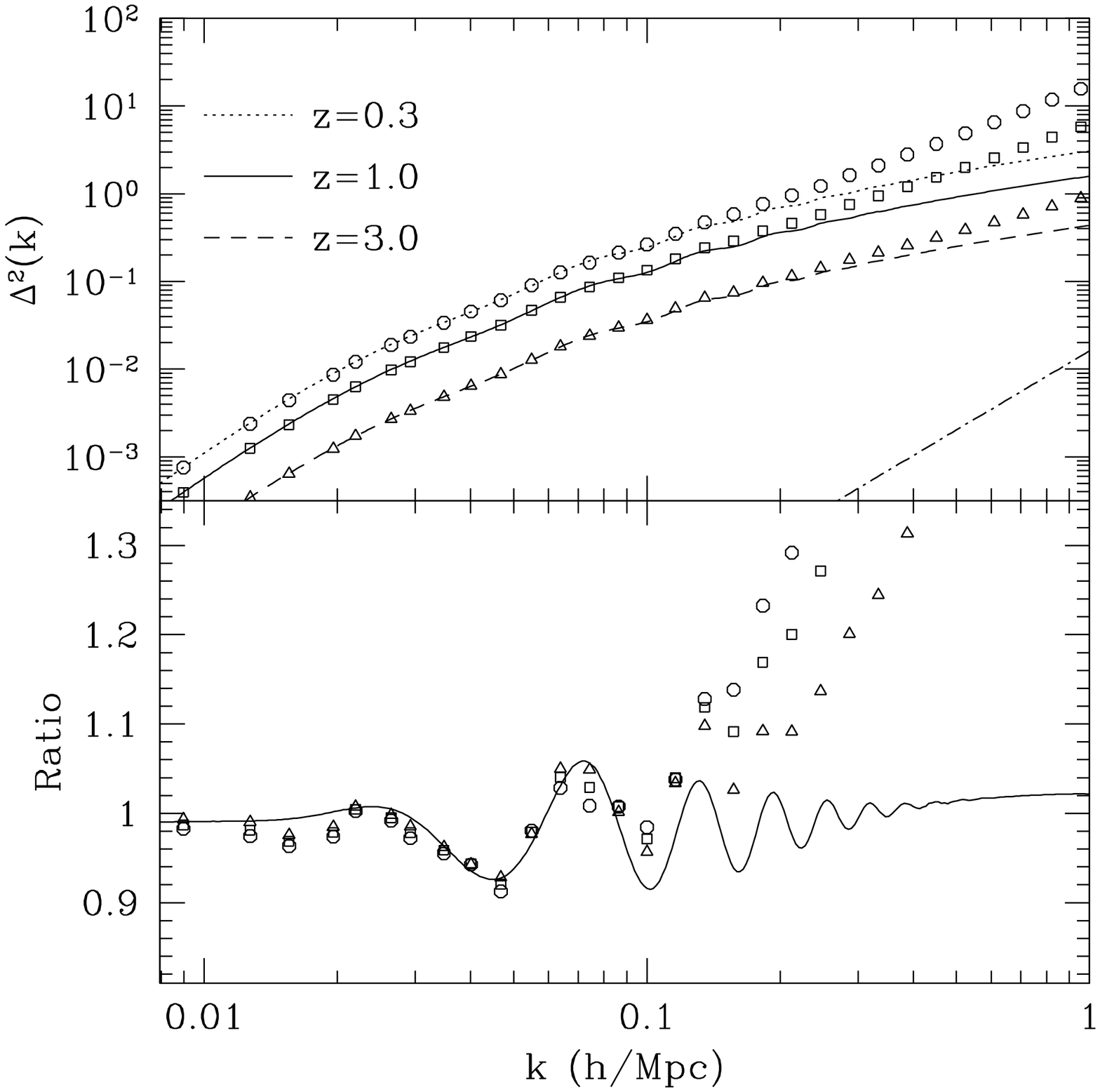,width=6.4in} 
\caption{Baryon acoustic oscillations mostly preserve their origin 
from primordial linear perturbations of the photon-baryon fluid.  
But as matter perturbations grow, nonlinear effects gradually dissolve 
the standard ruler scale.  The top panel shows the mass power per 
logarithmic $k$ mode at three redshifts from an N-body simulation, 
compared to the linear results.  The bottom panel shows the ratio 
relative to the ``no baryon'' linear case, with the curve including 
baryons.  The simulation points deviate from the clear oscillatory 
structure even when the nonlinearities are quite modest.  Figure 
courtesy of M.\ White \citep{mwhite05}. 
} 
\label{fig.nonlin}
\end{center} 
\end{figure}

Such a characterization of linearity, e.g.\ requiring the mass 
fluctuation amplitude to be less than 0.5, is likely overoptimistic. 
Figure \ref{fig.nonlin} shows the results of a $1024^3$ particle, 
$2048^3$ grid PM simulation in a 700 $h^{-1}$Mpc box, from M.\ White 
\citep{mwhite05}.  The curves in the upper panel show the linear theory 
prediction for the mass variance per logarithmic wavemode $k$ and the 
points give the simulation results, showing the deviation at 
higher $k$ as nonlinear effects enter.  The bottom panel shows 
the ratio of power relative to a zero baryon content universe, but 
the main point is the rapidity with which the simulation results 
(including nonlinear effects) deviate from the linear theory curve 
wherein the BAO are readily apparent -- even though the nonlinear 
effects are only at the 5-10\% level.  At $z=1$, one might be able 
to convince oneself one sees three peaks, and at $z=3$ maybe four, 
but there is a huge difference between being able to detect that 
a peak exists and being able to characterize the wavelength at the 
$\le$1\% level. 

Careful work needs to be carried out to determine how much residual 
uncertainty is caused by nonlinear mode coupling, and by lack of 
complete understanding of the true nonlinear mass power spectrum. 

\subsection{Redshift space distortions} 

Baryon acoustic oscillations show themselves in the three dimensional 
spatial pattern of the baryon distribution on large scales.  Besides 
the difficulties due to not detecting baryons directly (\S\ref{sec.bias}), 
complications arise due to not detecting the radial distance directly. 
This is the well known problem of translating from a redshift space 
distribution to a real space distribution.  In particular, massive 
structures show three effects: a stretching along the line of sight 
(finger of god) effect due to internal velocity dispersion in the 
gravitational potential, a squashing due to large scale infall, and 
a distortion due to the spacetime geometry 
shear effect discussed by \citet{ap} 
(basically arising because points at the same distance emit light 
signals at the same value of the cosmic expansion, while points 
with radial separation emit at different values). 

The formalism for dealing with these distortions has been developed 
in various limits, e.g.\ by \citet{kaiser,ballinger,matsu}.  
These distortions, needing to be fit, add some level of uncertainty 
to determining the baryon acoustic oscillation scale.  
One can employ a polynomial fit to the slope induced by this systematic 
in the power spectrum to reduce its effect.  With such a procedure, 
the redshift distortion clearly does not prevent detection of the 
peaks -- however we do not have quantification of how it affects 
the precision characterization of the peaks, i.e.\ is the scale 
still accurately determined at the $\le1$\% level?  Recall that 
we are interested in the $k$-space scale of the oscillations, so 
even if the residual amplitude of the systematics after correction 
is at a few percent 
this does not guarantee recovery of the length scale to $\le1$\%. 

Moreover, if we rely on the theory of redshift distortions as 
mentioned above, we must take into account the results by \citet{scocci} 
showing the failure of its elements outside their limiting validity, 
in particular the influence of nongaussianities and nonlinearities.  
It seems more robust to 
attempt to make such corrections through simulations.  This can certainly 
be done but will require a comprehensive suite of various cosmological 
models. 

\subsection{Other Systematics} 

Other systematics that enter at a low, but not yet defined, level 
include selection function effects.  As for SN, the fact that 
galaxies serve merely as markers for the BAO method means that we 
can pick and choose a sparse sample of objects.  This helps reduce 
the time and cost of a survey, but the sample cannot be too sparse 
or shot noise effects will enter once the survey has been divided 
into redshift bins or subclasses (for crosschecks).  Care must be 
taken that the selection is 
homogeneous in any quantity that contains cosmology dependence. 
In determining precision redshifts, one must avoid line confusion 
or blending, and be wary of the effect of intragalactic structure 
variation over a spectrograph slit, such as from star forming regions. 

BAO indeed have strong advantages 
in avoiding flux and color calibration issues (of course, if 
such calibration is accomplished for a supernova program, it will 
have widespread benefits for most fields of astronomy, but that is 
not a strictly dark energy issue).  Similarly, dust extinction should not be a 
problem, unless Milky Way extinction somehow apodizes power on 
large scales to confuse oscillations (which seems farfetched).  Similarly, 
as long as a BAO survey is properly designed so flux threshold limits 
do not interfere with redshift completeness, gravitational lensing 
magnification should not be an issue.  It could enter through distortion 
of the standard ruler -- remember lensing magnifies flux by magnifying 
scales, i.e.\ so-called convergence lensing or scale shrinking \citet{lin88a} 
-- but this should be negligible on the large scales involved in BAO. 

Some theory systematics exist as well.  One example already mentioned 
is the details of the nonlinear power spectrum: this probably does not 
have features harmonically related to the BAO scale, but incomplete 
correction for even ``broadband'' tilts can degrade determination of 
the scale.  On the standard ruler side of the calculation, the major 
uncertainty is the value of the physical matter density $\Omega_m h^2$; 
the Planck CMB survey should determine this to 0.9\%, which is precise but 
not perfectly so.  \citet{gb05} show that the residual 
uncertainty on the matter density translates into an up to 40\% 
degradation of the constraints on dark energy parameters.  
\citet{eiswh} showed that effects of curvature and neutrinos 
will not adversely affect the standard ruler use.  Computation of the 
CMB power spectrum, for fitting to the photon acoustic oscillations, 
is also not perfect: \citet{coracmb} showed that 10-15\% errors 
in CMBfast can occur for dynamical dark energy models, at both the 
lowest multipoles and in the acoustic peaks region. 

Perhaps most worrisome is that the matter power spectrum depends 
on the standard scenario in the dark matter and dark 
energy physics.  If we lose the gamble and 
the dark energy is not simple -- e.g.\ sound speed not equal to the 
speed of light, anisotropic stress nonzero, or coupling nonzero -- then 
these properties confuse the BAO technique.  For example they can change 
the turnover scale in the pure CDM power spectrum used to calibrate the 
oscillations and can even add new oscillations.  In this sense BAO are 
not a geometric probe, not following directly from the metric as SN 
distances do.

\subsection{Summary of Influence of Systematics} 

The method of BAO clearly has several major pluses in the area of 
systematics control, such as substantial freedom from flux calibration 
issues (though it does still have spectral, or redshift, calibration 
issues).   While the linearity of the physics is a strong plus, it 
is not total, and indeed the effects of nonlinearities might be the 
most worrying of those mentioned above.  Still, this should be less 
severe than for the 
weak lensing method.  All the issues brought up in this section (with 
the possible exception of the ``non-geometric'' one) are 
likely to be tractable with sufficient effort.  But that effort has 
not yet been put in, and so we cannot say for sure that all the 
systematic uncertainties will not contribute at the 1\% level. 

In the next section we will consider BAO as being purely statistically 
limited, without systematic errors, and see that the question of whether 
1\% accuracy can be achieved is crucial. 

\vspace{0.2in}

\section{BAO are more precise distance indicators than SN, {\it or} \\ 
SN are more precise dark energy indicators than BAO?} \label{sec.dist} 

If our goal is understanding dark energy, then we must keep this in 
mind when discussing the precision of measurements carried out for 
different techniques.  For example, the CMB power spectrum, and 
the distance to the last scattering surface, can be 
measured quite precisely, but they contain relatively little leverage 
on determining dark energy properties.  

One issue, related to the richness/meagerness argument above, is 
the density of distance measurements.  That is, if we are using distances 
to map the expansion history of the universe, how fine in detail is 
the map?  BAO are fundamentally limited in that millions of galaxies 
must be binned together to provide an accurate measurement of the 
oscillation wavelength, and crucially one cannot subdivide the redshift 
bins below this scale.  Since the scale corresponds to a comoving 
size of 100 $h^{-1}$ Mpc, the Nyquist frequency of the oscillations 
imposes a requirement that $\Delta z\ge0.2$.  SN, by contrast, can 
map the expansion arbitrarily finely, subject only to observational 
constraints not any fundamental limitation.  Note that such a difference in 
ability is not included in the following analysis in terms of 
a smooth, slowly varying equation of state. 

It is important to remember that any estimations of dark energy 
constraints must use a well behaved description of dark energy. 
Equation of state parametrizations that blow up quickly to 
unphysical values will give hypersensitive, inaccurate constraints. 
For example the parametrization $w(z)=w_0+w_1z$ can overstate 
the dark energy constraints by a factor 3 (\citet{eospar}). 

Calculations, e.g.\ \citet{seoeis03,seoeis05,gb03,gb05}, show that 
10000 square degree spectroscopic redshift surveys can achieve percent, or 
possibly subpercent, statistical precision on distances.  If the systematic 
uncertainties do not degrade this, it is comparable to or better than 
SN distance measurements, which are at the 1\% level.  Moreover, the 
radial modes of BAO provide the Hubble parameter $H(z)$, rather than 
the angular distance that is the integral of this quantity.  Since 
dark energy enters the distances through $H(z)$, a determination of 
the bare quantity seems advantageous.  These properties of BAO appear 
promising. 

However, we must then propagate the measured distances through to the 
dark energy constraints, and here a subtlety arises.  As stated in the 
Introduction, one actually measures relative distances, in both the 
BAO and SN cases.  The SN distances can be viewed as either luminosity 
distances to some redshift $z$ convolved with an additional parameter 
${\cal M}$ involving the absolute luminosity and Hubble constant, or 
equivalently the luminosity distance to some redshift $z$ relative to 
some low redshift value (formally $d(10\ pc)$).  The BAO distances 
involve the 
sound horizon scale, found through CMB angular scale measurements to 
the decoupling epoch (formally the last scattering surface).  So they 
can be thought of as distances to some redshift $z$ relative to the 
value at a high redshift, $z=1089$. 

Since dark energy is more dominant in the recent universe than in the 
high redshift universe, distances to $z=1.7$, say, measured relative 
to low redshifts are much more sensitive to dark energy properties 
than those measured relative to high redshifts.  Indeed, canonical 
models of dark energy have not merely subdominant energy densities at 
high redshift but basically negligible energy densities.  For example, 
a cosmological constant model has $\Omega_\Lambda(z=1.7)=0.1$ and 
$\Omega_\Lambda(z=1089)=10^{-9}$. 

So even if BAO could achieve somewhat more precise distance measurements 
than SN, plus the measurement of $H(z)$, SN could still achieve denser and 
more leveraging measurements.  This was calculated explicitly in 
\citet{linbo} and is examined in more detail in calculations for 
this article.  We find that high accuracy, relatively low redshift 
($z\approx0.5$), BAO measurements are important for dark energy 
leverage (note that here nonlinearities will be most severe).  
As expected, this is not as crucial for dynamical models 
that retain more dark energy at $z>1$.  

\subsection{Dark Energy Equation of State Constraints} \label{sec.de} 

As a baseline model, we consider 1\% measurements of both the radial 
and tangential BAO scale at redshifts $z=0.5$, 1, 2.75, 3.25.  We find 
that lack of data in the intermediate range $z\approx1.3-2.5$ (more 
difficult for ground based observations) does not change the broad 
characteristics.  Results are tested for a fiducial cosmological 
constant model and a SUGRA ($w_0=-0.82$, $w_a=0.58$) model, both 
with $\Omega_m=0.28$.  BAO 
is compared or added to other data sets such as the distance to CMB 
last scattering of Planck quality, SN distance measurements of SNAP 
quality (including systematics), or weak lensing (WL) shear power 
spectrum of SNAP 1000 square degrees quality. 

\begin{figure}[!hbt]
\begin{center} 
\psfig{file=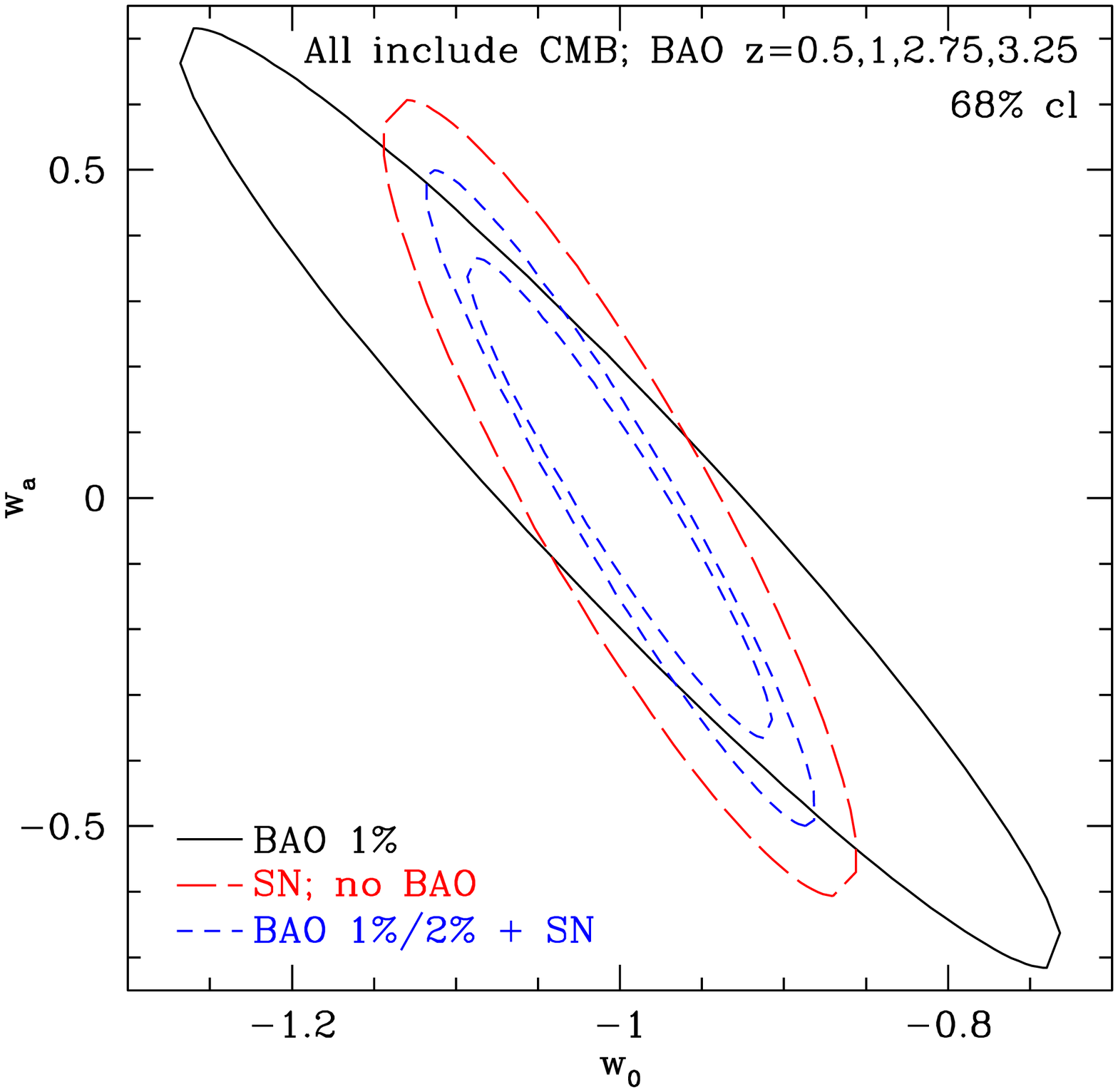,width=6.5in} 
\caption{Dark energy equation of state parameter estimates for 
various combinations of cosmological probes using next generation 
data.  Even 1\% BAO do not match the leverage of SN, but BAO can 
provide useful complementarity and a crosscheck. 
} 
\label{fig.distbosn}
\end{center} 
\end{figure}

We illustrate some results in Figure \ref{fig.distbosn}, including 
the effect of diluting the BAO measurement accuracy to 2\%.  
Parameters not shown, such as $\Omega_m$ or ${\cal M}$, are 
marginalized over.  Note that 
since nonlinear effects are most troublesome at the $z=0.5$ bin, this 
point, which carries substantial dark energy leverage, may be the hardest 
place to attain 1\% accuracy.  BAO+CMB yield 0.18, 0.47 one sigma uncertainty 
on $w_0$, $w_a$, while SN+CMB give 0.09, 0.37, and all three together 
provide 0.06, 0.24.  The complementarity is significant.  Note that 
only all three probes together begin to approach the ``revelatory'' 
requirement of estimating $w'=w_a/2$ to 0.1.  If BAO is 
weakened to 2\%, then BAO+CMB gives a poor 0.35, 0.90, and the 
trio provides 0.08, 0.33 -- minor improvement over the case without BAO. 
Thus, the accuracy attainable with the BAO method is important to 
know rigorously.

We have seen from the above results that SN data are a key element 
for the dark energy constraints (e.g.\ improving estimation of $w_0$ 
by a factor of 2-4 alone, and a factor 3-4 added to BAO).  But given the 
complementarity, if BAO can provide 1\% measurements then the SN data 
set may not need to be as stringent.  Reducing the survey depth to, 
say, $z=0.8$ but somehow keeping 
the SN systematics at the same low level as for the space based SNAP 
survey, degrades the BAO+CMB+SN constraints only to 0.06, 0.25. 
However if the BAO precision slips to 2\% this becomes 0.09, 0.42. 
Careful study is required.  We will also see later that the SN depth 
is an important element in several other respects. 

Note that just because BAO effectively involves a distance ratio of, 
say, $d(z=3)$ to $d(z=1089)$, this does not mean that BAO can simply 
separate out the conditions of the universe between $z=3$ and $z=1089$. 
That is, one does not isolate the effects of ``everything but'' dark 
energy or the effects of unexpected early dark energy.  Both distances 
entering the ratio are still integral quantities, and while the 
conditions at $z>3$ are involved, they are not given separately.  Still, 
BAO does offer the possibility of putting some constraints on $z>3$ 
dark energy (though one expects the mass growth factor to be more 
sensitive to this property).

\subsection{Complementarity with Cosmic Growth Probes} \label{sec.grow} 

None of the probes considered above have dependence on mass growth, 
and so are 
incapable of comparing the expansion history vs.\ the growth history 
to test the theoretical framework (see, e.g., \citet{groexp}).  
Whether the dark energy arises 
from a new physical component, e.g.\ a high energy physics scalar 
field, or a modification of the theory of gravity is a crucial 
question.  Answering this is a key requirement for understanding 
the physics of acceleration.  Thus, just as 
a purely SN experiment would not be sufficiently revelatory about 
dark energy, a purely BAO experiment is not acceptable.  
We therefore consider measurement of the weak lensing shear power 
spectrum, as estimated for the SNAP satellite, as another probe of 
dark energy to be taken in complementarity.  Note that this, like 
BAO but {\it unlike\/} SN, is here treated with purely statistical errors. 

\begin{figure}[!hbt]
\begin{center} 
\psfig{file=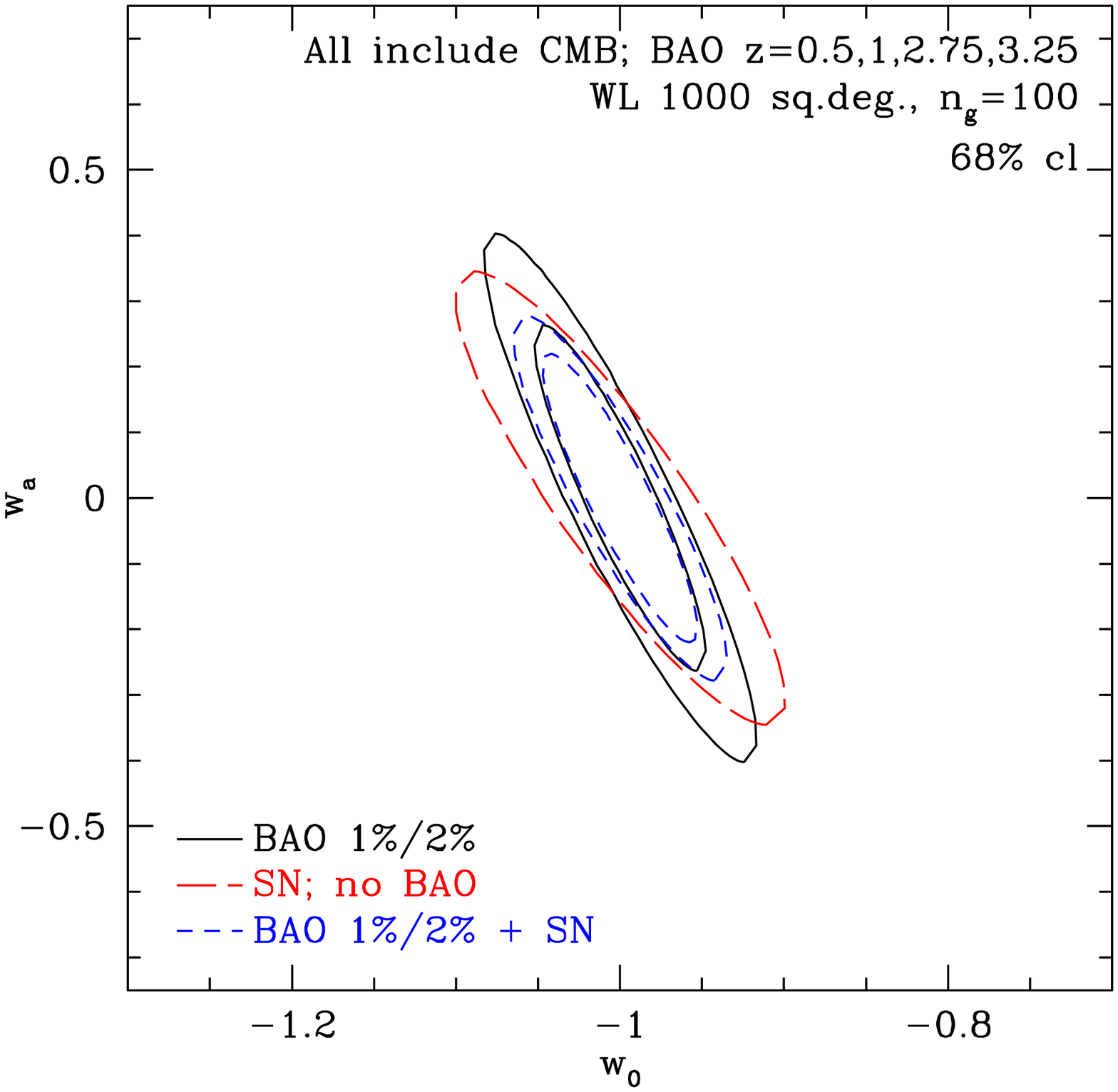,width=6.5in} 
\caption{Dark energy equation of state parameter estimates for 
various combinations of cosmological probes using next generation 
data.  Weak lensing complements both BAO and SN, and all probes 
together provide strong constraints. 
} 
\label{fig.bocs}
\end{center} 
\end{figure}

Figure \ref{fig.bocs} shows that WL adds appreciable information 
enabling tighter dark energy constraints.  Recall, however, that we also 
want each probe, or at least expansion history and growth history 
separately, to stand on their own to allow for crosschecks and 
test of the theoretical framework (e.g.\ modifications of Einstein 
gravity).  With this kept firmly in mind, we examine the constraints upon 
combining WL data with the other probes. 

First, note that WL+CMB without either BAO or SN gives constraints 
on $w_0$, $w_a$ of 0.13, 0.49, so complementarity with other probes 
is desired.  (One could also consider extension of the area of the 
space quality survey, or addition of other weak lensing techniques 
such as higher order correlations or cross-correlation cosmography. 
This would help the WL probe stand as more comparable to the expansion 
history constraints.) 

For SN+CMB+WL, the constraints are 0.066, 0.23; BAO+CMB+WL attain 
0.034 (0.055), 0.17 (0.27) for BAO at the 1\% (2\%) level.  Again we 
see that it is important to find whether the systematic uncertainties 
allow BAO determination at the 1\% level -- and to realize that SN 
are the only pure geometric probe, immune to microphysics in the dark 
energy sector.  All probes combined give 
dark energy bounds of 0.031, 0.14.  We find that BAO does add value 
to SN and WL methods, and SN adds value to BAO and WL methods, especially 
when the BAO accuracy is 2\%.  Reducing the SN survey depth to $z=0.8$ 
is more harmful now, with degradations in the case of all probes 
combined up to the 22\%, 39\% level on $w_0$, $w_a$.

Note that synergy exists between surveys carrying out WL and BAO 
measurements.  A large scale photometric 
survey for WL would basically supply a large scale photometric BAO 
survey and be an important selector for a large scale spectroscopic BAO 
survey, while a spectroscopic BAO survey can help calibrate photometric 
redshifts for the WL survey.  Increasing the area of the WL survey to 
4000 square degrees provides 20\% improvement in the dark energy 
parameter constraints using all the probes.  Keeping WL at 1000 square 
degrees but adopting the dynamical SUGRA fiducial model gives complete 
combination 
constraints of $\sigma(w_0)=0.020$, $\sigma(w'=w_a/2)=0.029$, with a 
minimum variance equation of state uncertainty $\sigma(w_{\rm min}= 
w(z=0.43))=0.0095$! 

This combination of probes seems quite exciting in their prospects for 
revealing the nature of dark energy -- if the systematic uncertainties 
can be kept below the statistical levels employed here.  Also remember 
that we should require tight constraints not only from all probes jointly, 
but from each individually.  Only thus will we have confidence in the 
new physics discovered. 

\subsection{Other Cosmological Probe Aspects of BAO} \label{sec.other} 

One can consider other uses of BAO measurements, such as constraints 
on spatial curvature in conjunction with the cross-correlation 
cosmography technique of weak lensing (\citet{bernomk}).  Note 
that this aims at a 
measurement of curvature, not a test of the constancy of the spatial 
curvature over cosmic time.  The latter is sometimes mistakenly attributed 
to CMB measurements, saying that the CMB measures curvature at the 
epoch of last scattering and this can be compared to measurements today. 
This statement is untrue, since the CMB measures the integral of 
the curvature effects (insofar as they can be separated from all the 
other component energy densities) from $z=0-1089$. 

While measuring the spatial curvature is a worthy goal, there are 
some aspects to consider.  It is not directly telling us anything about 
the nature of dark energy, though it does help 
through breaking degeneracies.  Furthermore, the proposed measurement 
method relies on two techniques, neither of which has been matured 
through a history of implementation and systematics studies.  So one 
should not use this to drive ``optimization'' of other probes 
around these (if that were even possible).  On the other hand, 
SN+WL also measures curvature to 1-2\%.  Finally, 
measurements at moderately high redshifts, e.g.\ around $z\approx3$ that is 
one of the two main ranges for applying BAO, are actually extraordinarily 
insensitive to spatial curvature: the dependence of cosmological 
distances on curvature goes through a null at $z\approx3$ (see \citet{curv} 
for further examination of curvature). 

One can also consider BAO as a means of testing the reciprocity, or 
thermodynamic, relation.  This is phrased either as a redshift scaling 
between angular distances and luminosity distances or ``third party'' 
angular distances between points not including the observer.  Most 
commonly the relation is phrased as $d_l=(1+z)^2 d_a$.  This has a 
long history in cosmology, dating from the 1930s with Tolman, Ruse, 
Etherington, through \citet{wbg}, and a general proof in 
terms of the Raychaudhuri equation and the second law of thermodynamics 
by \citet{lin88b}.  Due to this thermodynamic origin (basically, if 
two identical blackbodies sent photons to each other over cosmic distances 
then work would be done unless the relation held), it is of very general 
applicability.  As long as the propagating photons obey Liouville's 
Theorem, then the relation must work.  In this sense it is an equivalent 
problem to measuring the evolution of the CMB temperature: if 
$T(z)\ne T_0(1+z)$, 
we would far more likely blame the measurement than believe a breakdown 
of physics.  So using BAO (measuring $d_a$) and SN (measuring $d_l$) 
to test for violation of the reciprocity relation is not likely to be 
actually useful (apart from the slim chance of there really being a 
violation such as from photon-axion mixing, say).  It is an interesting 
idea, but one would not advocate having $T(z)$ measurements drive CMB 
observations either. 

\section{Conclusion} 

In summary, baryon acoustic oscillations offer another promising cosmological 
probe.  Astrophysicists should certainly pursue its development, 
theoretically, algorithmically, and observationally, to learn how to 
practically carry out large spectroscopic galaxy redshift surveys and 
extract the information, and to obtain realistic 
estimation of its accuracy.  Ground based observations 
serve as the starting point, and probably dominant source, of data. 
A 10000 square degree spectroscopic survey, free from systematics, is 
a challenging endeavor.  
Space observations may have a role to play in covering the redshift 
range $z\approx1-2$, though this does not appear to be crucial, but 
the reduced noise from space could play a useful role. 

It is important to keep in mind, 
that BAO is not a panacea nor is it effortless.  Corrections do need 
to be applied, and the residual systematics, while promising, require 
hard work to quantify at the 1\% level.  
Recall that BAO is intrinsically limited 
in the fineness of the expansion history mapping possible, and less 
sensitive to dark energy for the same precision due to its tie to 
the matter dominated era.  Finally, a ``nonstandard'' dark matter or 
dark energy sector could throw BAO awry since they are not a purely 
geometric probe, while SN remain clean. 

The presence of such ups and downs holds for any cosmological probe. 
Given the few ways we have of robustly understanding the new physics 
behind the accelerating universe, the baryon acoustic oscillations method 
is a welcome addition.  Pressing forward with further study, and actual 
observational application, BAO may offer 
important complementarity to supernovae, and weak lensing, probes 
for understanding dark energy.  

\begin{acknowledgments} 
This work has been supported in part by the Director, Office of Science,
Department of Energy under grant DE-AC02-05CH11231.  I thank Chris Blake, 
Daniel Eisenstein, Gary Bernstein, Karl Glazebrook, and Martin White for 
useful conversations. 
\end{acknowledgments}


\begin{thebibliography}{99} 

\bibitem[Alcock \& Paczy{\'n}ski(1979)]{ap} 
C.\ Alcock \& B.\ Paczy{\'n}ski 1979, Nature 281, 358 

\bibitem[Ballinger, Peacock, \& Heavens(1996)]{ballinger} 
W.E.\ Ballinger, J.A.\ Peacock, \& A.F.\ Heavens 1996, MNRAS 282, 877 

\bibitem[Bernstein(2005)]{bernomk} 
G.\ Bernstein 2005, astro-ph/0503276 

\bibitem[Blake \& Glazebrook(2003)]{gb03} 
C.\ Blake \& K.\ Glazebrook 2003, Ap.J.\ 594, 665 

\bibitem[Corasaniti et al.(2004)]{coracmb} 
P-S.\ Corasaniti et al.\ 2004, Phys.\ Rev.\ D 70, 083006 

\bibitem[Dolney et al.(2004)]{dolney} 
D.\ Dolney, B.\ Jain, \& M.\ Takada 2004, MNRAS 352, 1019 

\bibitem[Eisenstein et al.(2005)]{eis05} 
D.J.\ Eisenstein et al.\ 2005, Ap.J.\ 633, 560

\bibitem[Eisenstein(2005)]{eisanim} 
Eisenstein -- http://cmb.as.arizona.edu/~eisenste/acousticpeak/acoustic\_physics.html 

\bibitem[Eisenstein \& White(2004)]{eiswh} 
D.J.\ Eisenstein \& M.\ White 2004, Phys.\ Rev.\ D 70, 103523 

\bibitem[Glazebrook \& Blake(2005)]{gb05} 
K.\ Glazebrook \& C.\ Blake 2005, Ap.J.\ 631, 1 

\bibitem[Kaiser(1987)]{kaiser} 
N.\ Kaiser 1987, MNRAS 227, 1

\bibitem[Linder(1988a)]{lin88a} 
E.V.\ Linder 1988a, A\&A 206, 199 

\bibitem[Linder(1988b)]{lin88b} 
E.V.\ Linder 1988b, MPA internal research note 

\bibitem[Linder(2003)]{linbo} 
E.V.\ Linder 2003, Phys.\ Rev.\ D 68, 083504 

\bibitem[Linder(2005a)]{groexp} 
E.V.\ Linder 2005a, Phys.\ Rev.\ D 72, 043529

\bibitem[Linder(2005b)]{curv} 
E.V.\ Linder 2005b, Astropart.\ Phys.\ 24, 391

\bibitem[Linder \& Huterer(2005)]{eospar} 
E.V.\ Linder \& D.\ Huterer 2005, Phys.\ Rev.\ D 72, 043509 

\bibitem[Matsubara \& Szalay(2002)]{matsu} 
T.\ Matsubara \& A.S.\ Szalay 2002, Ap.J.\ 574, 1

\bibitem[Meiksin, White, \& Peacock(1999)]{mwp} 
A.\ Meiksin, M.\ White, J.A.\ Peacock, MNRAS 304, 851 (1999) 

\bibitem[Peebles \& Yu(1970)]{peeblesyu} 
P.J.E.\ Peebles \& J.T.\ Yu 1970, Ap.J.\ 162, 815 

\bibitem[Sakharov(1965)]{sakharov} 
A.\ Sakharov 1965, Soviet Physics JETP 22, 241 

\bibitem[Scoccimarro(2004)]{scocci} 
R.\ Scoccimarro 2004, Phys.\ Rev.\ D 70, 083007 

\bibitem[Seo \& Eisenstein(2003)]{seoeis03}
H-J.\ Seo \& D.J.\ Eisenstein 2003, Ap.J.\ 598, 720 

\bibitem[Seo \& Eisenstein(2005)]{seoeis05} 
H-J.\ Seo \& D.J.\ Eisenstein 2005, Ap.J.\ 633, 575

\bibitem[Sunyaev \& Zel'dovich(1970)]{sunyaev} 
R.A.\ Sunyaev  \& Ya.B.\ Zel'dovich 1970, Ap.\ \& Sp.\ Sci.\ 7, 20 

\bibitem[Weinberg(1972)]{wbg} 
S.\ Weinberg, Gravitation and Cosmology (Wiley, New York, 1972) 

\bibitem[White(2005)]{mwhite05} 
M.\ White 2005, Astropart.\ Phys.\ 24, 334

\end{thebibliography}
\end{document}